\documentclass[aps,preprint,amsmath,amssymb,11pt]{revtex4}
\usepackage{graphicx}
\usepackage{epstopdf}
\pdfoutput=1 
\usepackage[T1]{fontenc}
\newcommand\sss{\scriptscriptstyle}
\newcommand{\mW}{m_{\sss W}}
\newcommand{\sW}{s_{\sss W}}
\newcommand{\cW}{c_{\sss W}}

\begin{document}

\title{Testing for observability of Higgs effective couplings in triphoton production at FCC-hh}
\author{H. Denizli}
\email[]{denizli_h@ibu.edu.tr}
\author{K. Y.  Oyulmaz}
\email[]{kaan.oyulmaz@gmail.com} 
\author{A. Senol}
\email[]{senol_a@ibu.edu.tr} 
\affiliation{Department of Physics, Abant Izzet Baysal University, 14280, Bolu, Turkey}
\begin{abstract}

We investigate the potential of the $pp\to \gamma\gamma\gamma$ process to probe CP-conserving and CP-violating dimension-six operators of Higgs-gauge boson interactions in a model-independent Standard Model effective field theory framework at the center of mass energy of 100 TeV which is designed for the Future Circular hadron-hadron Collider. Signal events assuming the existence of anomalous Higgs boson couplings at $H\gamma\gamma$ and $HZ\gamma$ vertices and the relevant SM background events are generated in MadGraph, then passed through Pythia 8 for parton showering and Delphes to include detector effects. After detailed examination of kinematic variables, we use the invariant mass distribution of the two leading photons with optimized kinematic cuts to obtain constraints on the Wilson coefficients of dimension-six operators. We report that limits at 95\% confidence level on  $\bar{c}_{\gamma}$ and  $\tilde{c}_{\gamma}$ couplings with an integrated luminosity of 10 ab$^{-1}$ are [-0.0051; 0.0038]  and [-0.0043; 0.0043] without systematic error, respectively.


\end{abstract}

\maketitle

\section{Introduction}
The investigation of the Higgs sector of Standard Model (SM) responsible for the mechanism of the electroweak symmetry breaking has become an attraction point in particle physics after the ATLAS and CMS collaboration's discovery of a scalar particle with a mass of 125 GeV which is compatible with the predicted Standard Model (SM) Higgs boson \cite{Aad:2012tfa, Chatrchyan:2012xdj}. Thus, the precision measurements of the Higgs couplings have a great potential to shed light on the new physics beyond the SM involving massive particles that are decoupled at energy scales much larger than the Higgs sector energies. One of the well-known investigation methods looking for a deviation from SM is the Effective Field Theory (EFT) approach which is based on potential new physics contributions beyond the SM effects described by a systematic expansion in a series of high dimensional operators beyond the SM fields as well as SM operators \cite{Buchmuller:1985jz,Grzadkowski:2010es}. Since the dimension-6 operators are loop induced, their matching to ultraviolet (UV) models is simplified by universal one-loop effective action \cite{Appelquist:1974tg}. Therefore, the dimension-6 operators play an important role in the EFT framework. There have been many studies on EFT operators between Higgs and SM gauge boson via different production mechanisms at hadron colliders \cite{Hagiwara:1993qt,Corbett:2012ja,Ellis:2014jta,Ellis:2014dva,Falkowski:2015fla,Corbett:2015ksa,Ferreira:2016jea,Aad:2015tna,Englert:2015hrx,Buckley:2015lku,Khanpour:2017inb,Englert:2016hvy,Degrande:2016dqg,Bishara:2016kjn,Liu-Sheng:2017pxk,Aaboud:2018xdt,Khachatryan:2014kca,Khachatryan:2016tnr}. Among the production mechanisms in the hadron colliders, events containing three photons in the final state provides an ideal platform to search for deviations from SM since it is rare in the SM and involves only pure electroweak interaction contributions at tree level \cite{deCampos:1998xx,GonzalezGarcia:1999fq,Aaboud:2017lxm}. In addition, prior to these searches at hadron colliders, there were similar searches for anomalous Higgs couplings using EFT formalism in the triphoton final state (and others) as for example \cite{Achard:2004kn,Abreu:1999vt,Heister:2002ub}. The triphoton can be produced in hadron-hadron collisions either in the hard interaction via annihilation of an initial state quark-antiquark pair  which is called direct production or from the  fragmentation of high $p_T$ parton which is called fragmentation process. Since photons produced via direct production are typically isolated, requiring isolated photons will reduce the background contributions from the decays  of unstable particles such as $\pi^0\to \gamma\gamma$ and suppress the signal process with one or more fragmentation photons.

One of the future  projects currently under consideration by CERN is the Future Circular Collider (FCC) facility which would be built in a 100 km tunnel and designed to deliver $pp$, $e^+e^-$ and $ep$ collisions \cite{fcc}. The FCC facility which has the potential to search for a wide parameter range of new physics is the energy frontier collider project following the completion of the LHC and High-luminosity LHC physics programmes. One of the FCC options, the FCC-hh, is designed to provide proton-proton collisions at the proposed 100 TeV centre-of-mass energy with peak luminosity $5\times10^{34}$ $cm^{-2}s^{-1}$ \cite{fcchh,Mangano:2017tke}. 

In this study, we investigate the potential of the process $pp\to\gamma\gamma\gamma$ at FCC-hh in the existence of anomalous Higgs boson couplings at $H\gamma\gamma$ and $HZ\gamma$ vertices. Description of the SM EFT Lagrangian is given in the next section. Details of the analysis including event generation, detector effects and event selection as well as statistical method used to obtain the limits on the anomalous Higgs-neutral gauge boson couplings are illustrated in section III. Our results for an integrated luminosity of 10 ab$^{-1}$ are presented and discussed in the last section.

\section{Effective Operators}
The most general form of effective Lagrangian including dimension-6 operators expressed in the basis convention of the Strongly Interacting Light Higgs (SILH)  \cite{Alloul:2013naa} as well as SM is given as follows;
\begin{eqnarray}
\mathcal{L}_{eff} = \mathcal{L}_{\rm SM} + \sum_{i}\bar c_{i}O_{i}+\sum_{i} \tilde c_{i}O_{i}
\end{eqnarray}
where  $\bar c_{i}$ and $\tilde c_{i}$ are normalized Wilson coefficients of the CP-conserving and CP-violating interactions, respectively. In this work, we focused on the CP-conserving and CP-violating interactions of the Higgs boson and electroweak gauge boson in the SILH basis.
The CP-conserving part of an effective Lagrangian is 
\begin{eqnarray}\label{CPC}
	\begin{split}
		\mathcal{L}_{\rm CPC} = & \
		 \frac{\bar c_{H}}{2 v^2} \partial^\mu\big[\Phi^\dag \Phi\big] \partial_\mu \big[ \Phi^\dagger \Phi \big]
		+ \frac{\bar c_{T}}{2 v^2} \big[ \Phi^\dag {\overleftrightarrow{D}}^\mu \Phi \big] \big[ \Phi^\dag {\overleftrightarrow{D}}_\mu \Phi \big] - \frac{\bar c_{6} \lambda}{v^2} \big[\Phi^\dag \Phi \big]^3
		\\
		& \
		  - \bigg[\frac{\bar c_{u}}{v^2} y_u \Phi^\dag \Phi\ \Phi^\dag\cdot{\bar Q}_L u_R
		  + \frac{\bar c_{d}}{v^2} y_d \Phi^\dag \Phi\ \Phi {\bar Q}_L d_R
		+\frac{\bar c_{l}}{v^2} y_l \Phi^\dag \Phi\ \Phi {\bar L}_L e_R
		 + {\rm h.c.} \bigg]
		\\
		&\
		 + \frac{i g\ \bar  c_{W}}{m_{W}^2} \big[ \Phi^\dag T_{2k} \overleftrightarrow{D}^\mu \Phi \big]  D^\nu  W_{\mu \nu}^k + \frac{i g'\ \bar c_{B}}{2 m_{W}^2} \big[\Phi^\dag \overleftrightarrow{D}^\mu \Phi \big] \partial^\nu  B_{\mu \nu} \\
		&\   
		+ \frac{2 i g\ \bar c_{HW}}{m_{W}^2} \big[D^\mu \Phi^\dag T_{2k} D^\nu \Phi\big] W_{\mu \nu}^k  
		+ \frac{i g'\ \bar c_{HB}}{m_{W}^2}  \big[D^\mu \Phi^\dag D^\nu \Phi\big] B_{\mu \nu}   \\
		&\
		 +\frac{g'^2\ \bar c_{\gamma}}{m_{W}^2} \Phi^\dag \Phi B_{\mu\nu} B^{\mu\nu}  
		+\frac{g_s^2\ \bar c_{g}}{m_{W}^2} \Phi^\dag \Phi G_{\mu\nu}^a G_a^{\mu\nu} 
	\end{split}
\end{eqnarray}
where $\Phi$ is the Higgs sector containing a single $SU(2)_L$ doublet of fields; $\lambda$ is the Higgs quartic coupling; $g'$, $g$ and $g_s$  are coupling constant of  $U(1)_Y$, $SU(2)_L$ and $SU(3)_C$ gauge fields, respectively;  $y_u$, $y_d$ and $y_l$ are the $3\times3$ Yukawa coupling matrices in flavor space; the generators of $SU(2)_L$ in the fundamental representation are given by $T_{2k}=\sigma_k/2$ (here $\sigma_k$ are the Pauli matrices); $\overleftrightarrow{D}_\mu$ correspond to the Hermitian derivative operators; $B^{\mu\nu}$, $W^{\mu \nu}$ and $G^{\mu\nu}$ are the electroweak and the strong field strength tensors, respectively.  

 The effective Lagrangian in the SILH basis can be expanded to involve the extra $CP$-violating operators defined as,
\begin{eqnarray}
\label{CPV}
  {\cal L}_{CPV} = &\
    \frac{i g\ \tilde c_{ HW}}{\mW^2}  D^\mu \Phi^\dag T_{2k} D^\nu \Phi {\widetilde W}_{\mu \nu}^k
  + \frac{i g'\ \tilde c_{ HB}}{\mW^2} D^\mu \Phi^\dag D^\nu \Phi {\widetilde B}_{\mu \nu}
  + \frac{g'^2\  \tilde c_{ \gamma}}{\mW^2} \Phi^\dag \Phi B_{\mu\nu} {\widetilde B}^{\mu\nu}\\
 &\
  +\!  \frac{g_s^2\ \tilde c_{ g}}{\mW^2}      \Phi^\dag \Phi G_{\mu\nu}^a {\widetilde G}^{\mu\nu}_a
  \!+\!  \frac{g^3\ \tilde c_{3W}}{\mW^2} \epsilon_{ijk} W_{\mu\nu}^i W^\nu{}^j_\rho {\widetilde W}^{\rho\mu k}
  \!+\!  \frac{g_s^3\ \tilde c_{ 3G}}{\mW^2} f_{abc} G_{\mu\nu}^a G^\nu{}^b_\rho {\widetilde G}^{\rho\mu c} \ \nonumber
\end{eqnarray}
where \begin{eqnarray}
  \widetilde B_{\mu\nu} = \frac12 \epsilon_{\mu\nu\rho\sigma} B^{\rho\sigma} \ , \quad
  \widetilde W_{\mu\nu}^k = \frac12 \epsilon_{\mu\nu\rho\sigma} W^{\rho\sigma k} \ , \quad
  \widetilde G_{\mu\nu}^a = \frac12 \epsilon_{\mu\nu\rho\sigma} G^{\rho\sigma a} \ \nonumber 
\end{eqnarray}
are the dual field strength tensors.

The SILH bases of CP-conserving and CP-violating dimension-6 operators given in Eq.\ref{CPC} and Eq.\ref{CPV} can be defined in terms of the mass eigenstates after electroweak symmetry breaking. The Lagrangian with the relevant subset of anomalous Higgs and neutral gauge boson couplings in the mass basis for triphoton production is as follows
\begin{eqnarray}\label{massb}
  {\cal L} &= &\ 
    - \frac{1}{4} g_{\sss h\gamma\gamma} F_{\mu\nu} F^{\mu\nu} h
    - \frac{1}{4} \tilde g_{\sss h\gamma\gamma} F_{\mu\nu} \tilde F^{\mu\nu} h
\nonumber\\
    &-& \frac{1}{4} g_{\sss hzz}^{(1)} Z_{\mu\nu} Z^{\mu\nu} h
    - g_{\sss hzz}^{(2)} Z_\nu \partial_\mu Z^{\mu\nu} h
    + \frac{1}{2} g_{\sss hzz}^{(3)} Z_\mu Z^\mu h
    - \frac{1}{4} \tilde g_{\sss hzz} Z_{\mu\nu} \tilde Z^{\mu\nu} h
\\ 
    &-& \frac{1}{2} g_{\sss haz}^{(1)} Z_{\mu\nu} F^{\mu\nu} h
    - \frac{1}{2} \tilde g_{\sss haz} Z_{\mu\nu} \tilde F^{\mu\nu} h
    - g_{\sss haz}^{(2)} Z_\nu \partial_\mu F^{\mu\nu} h 
    \nonumber
\end{eqnarray}
where $Z_{\mu\nu}$ and $F_{\mu\nu}$ are the field strength tensors of $Z$-boson and photon, respectively. The effective couplings in the gauge basis defined as dimension-6 operators are given in Table~\ref{mtable} in which $a_{H}$ coupling is the SM contribution to the $H\gamma\gamma$ vertex at loop level.

\begin{table}[h]
\caption{The relations between Lagrangian  parameters in the mass basis (Eq.\ref{massb}) and the Lagrangian in gauge  basis (Eqs. \ref{CPC} and \ref{CPV}). ($c_W\equiv\cos \theta_W$, $s_W\equiv\sin \theta_W$)}  
\begin{ruledtabular}\label{mtable}
\begin{tabular}{ll}
  $g_{h\gamma\gamma}$= $a_{ H} - \frac{8 g \bar c_{ \gamma} \sW^2}{\mW}$ & $\tilde g_{ h\gamma\gamma}$  
     $= -\frac{8 g \tilde c_{ \gamma} \sW^2}{\mW}$ \\
    $g^{(1)}_{ hzz}$= $\frac{2 g}{\cW^2 \mW} \Big[ \bar c_{HB} \sW^2 - 4 \bar c_{ \gamma} \sW^4 + \cW^2 \bar c_{ HW}\Big]$& $g^{(2)}_{ hzz}$= $\frac{g}{\cW^2 \mW} \Big[(\bar c_{ HW} +\bar c_{ W}) \cW^2  + (\bar c_{ B} + \bar c_{ HB}) \sW^2 \Big]$ \\
    $g^{(3)}_{hzz}$=  $\frac{g \mW}{\cW^2} \Big[ 1 -\frac12 \bar c_{H} - 2 \bar c_{T} +8 \bar c_{\gamma} \frac{\sW^4}{\cW^2}  \Big]$& $\tilde g_{ hzz}$ =$\frac{2 g}{\cW^2 \mW} \Big[ \tilde c_{ HB} \sW^2 - 4 \tilde c_{ \gamma} \sW^4 + \cW^2 \tilde c_{ HW}\Big]$ \\
     $g^{(1)}_{ h\gamma z}$= $\frac{g \sW}{\cW \mW} \Big[  \bar c_{ HW} - \bar c_{HB} + 8 \bar c_{ \gamma} \sW^2\Big]$& $\tilde g_{ h\gamma z}$ =$\frac{g \sW}{\cW \mW} \Big[  \tilde c_{HW} - \tilde c_{ HB} + 8 \tilde c_{ \gamma} \sW^2\Big]$ \\
    $g^{(2)}_{h\gamma z}$= $\frac{g \sW}{\cW \mW} \Big[  \bar c_{HW} - \bar c_{ HB} - \bar c_{ B} + \bar c_{ W}\Big]$&   \\
     \end{tabular}
\end{ruledtabular}
\end{table}


This parametrization \cite{Alloul:2013naa} based on the formulation  \cite{Contino:2013kra} is not complete \cite{Alonso:2013hga,Brivio:2017bnu}  since it chooses to remove two fermionic invariants while retaining all the bosonic operators. However, this choice assumes completely unbroken U(3) flavor symmetry of the UV theory and flavor diagonal dimension-six effects. At the end, we only claim a sensitivity study for $\bar{c}_{HW}$, $\bar{c}_{HB}$, $\bar{c}_{\gamma}$, $\tilde{c}_{HW}$, $\tilde{c}_{HB}$ and $\tilde{c}_{\gamma}$ couplings and do not consider higher order electroweak effects.

\begin{figure}
\includegraphics[scale=1.0]{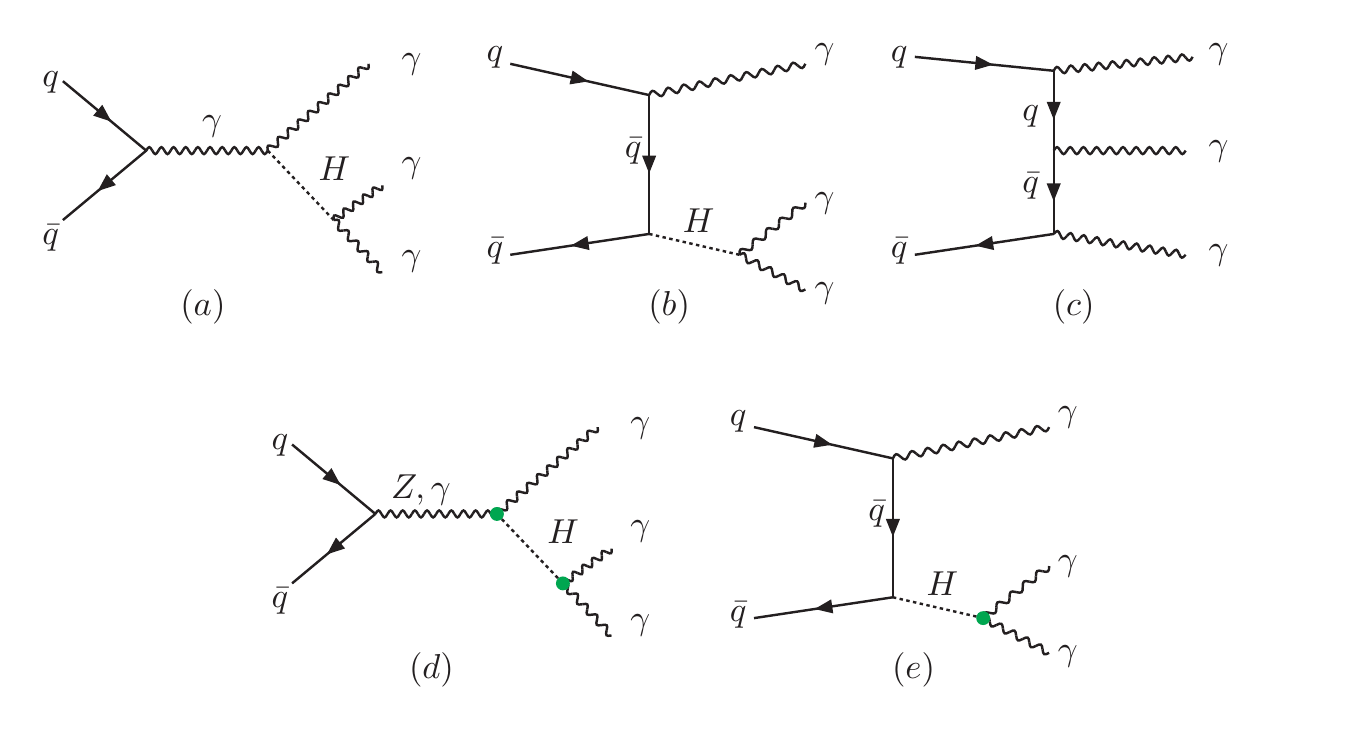}
\caption{The Feynman diagrams for ((a)-(c)) for SM and ((d)-(e)) the signal contribution with the effective Lagrangian of $q \bar q\to\gamma\gamma\gamma$ partonic subprocesses. \label{fd}}
\end{figure}

\begin{figure}
\includegraphics[scale=0.6]{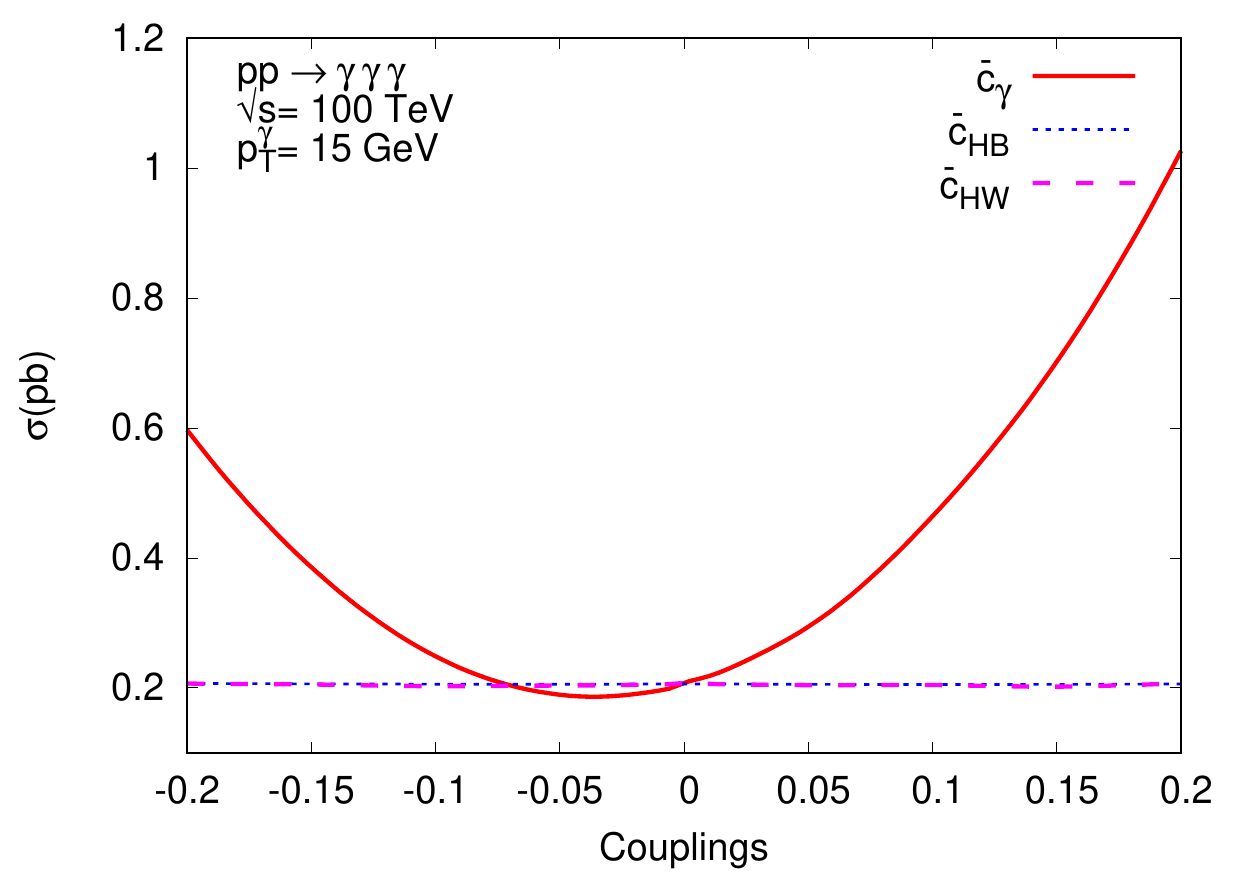}
\includegraphics[scale=0.6]{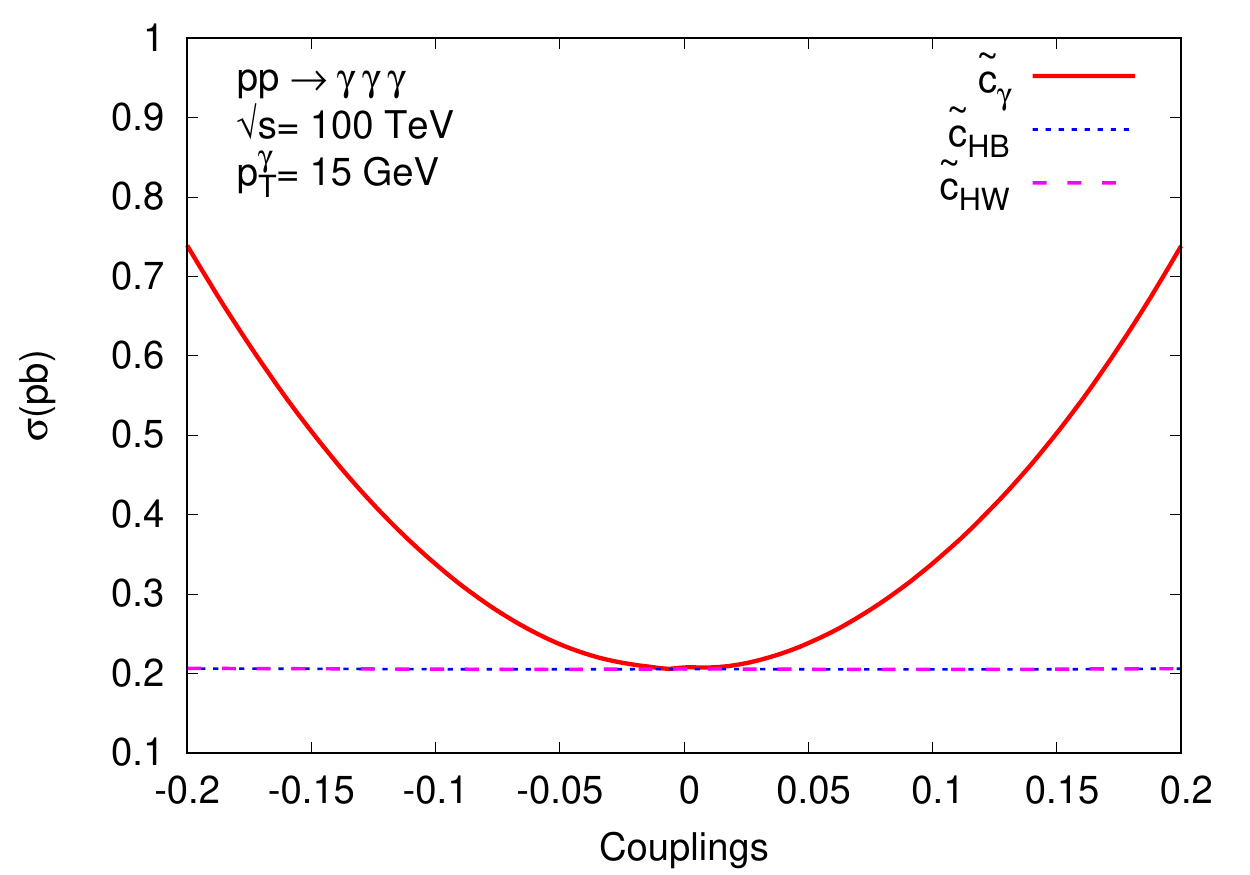}  
\caption{The total cross section as a function of the CP-conserving  $\bar{c}_{HW}$, $\bar{c}_{HB}$ and $\bar{c}_{\gamma}$ couplings (left) and CP-violating $\tilde{c}_{HW}$, $\tilde{c}_{HB}$ and $\tilde{c}_{\gamma}$ couplings (right) for $qq\to \gamma\gamma\gamma$ subprocess at the FCC-hh with $\sqrt s$=100 TeV. \label{crosssection}}
\end{figure}

Our study is based on the Monte Carlo simulations with leading order in \verb|MadGraph5_aMC@NLO v2.6.3.2| \cite{Alwall:2014hca} involving the effect of the dimension-6 operators on the triphoton production mechanism in $pp$ collisions. The effective Lagrangian of the SM EFT in Eq.(\ref{massb})  is implemented into the \verb|MadGraph5_aMC@NLO| event generator using FeynRules \cite{Alloul:2013bka} and UFO \cite{Degrande:2011ua} framework. The triphoton process is sensitive to Higgs-gauge boson couplings; $g_{h\gamma\gamma}$ and $g_{hz\gamma}$,  and the couplings of a quark pair to single Higgs field; $\tilde y_u$, $\tilde y_d$ in the mass basis.  On the other hand, this process is sensitive to the eight Wilson coefficients in the gauge basis:  $\bar c_{ W}$,  $\bar c_{ B}$, $\bar c_{ HW}$, $\bar c_{ HB}$, $\bar c_{ \gamma}$, $\tilde{c}_{HW}$, $\tilde{c}_{HB}$ and $\tilde{c}_{\gamma}$  related to Higgs-gauge boson couplings and also effective fermionic couplings. Due to the small Yukawa couplings of the first and second generation fermions, we neglect the effective fermionic couplings. We set $\bar c_{ W} + \bar c_{ B}$  to zero in all our calculations since the linear combination of  $\bar c_{ W} + \bar c_{ B}$  is strongly constrained from the electroweak precision test of the oblique parameters $S$ and $T$. 
The tree level Feynman diagrams of the leading-order partonic subprocesses for direct production ($q\bar q\to\gamma\gamma\gamma$) of three photon in pp collision are given Fig.\ref{fd}. The first three diagrams account for only SM background process (Fig.\ref{fd} (a)-(c)) while the others (Fig.\ref{fd} (d)-(e)) for the signal processes including anomalous $H\gamma\gamma$ and $H\gamma Z$ vertices. Fig.\ref{crosssection} shows the cross sections of $q\bar q\to\gamma\gamma\gamma$ partonic subprocess as a function of the CP-conserving $\bar{c}_{HW}$, $\bar{c}_{HB}$, $\bar{c}_{\gamma}$ couplings on the left panel and the CP-violating  $\tilde{c}_{HW}$, $\tilde{c}_{HB}$ and $\tilde{c}_{\gamma}$  couplings on the right panel. The cross sections are calculated at leading order including the transverse momentum( $p_T^{\gamma}$) and pseudo-rapidity ($\eta^{\gamma}$) cuts of all photons as $p_T^{\gamma} > 15$ GeV and $|\eta^{\gamma_{1,2}}| < 2.5$, respectively.  In this figure, one of the effective couplings is non-zero at a time, while the other couplings are fixed to zero and the SM cross section is recovered at the points $\bar{c}_{HW}$=$\bar{c}_{HB}$=$\bar{c}_{\gamma}$=0 and $\tilde{c}_{HW}$=$\tilde{c}_{HB}$=$\tilde{c}_{\gamma}$=0. One can easily see the deviation from SM for $\bar{c}_{\gamma}$ and $\tilde{c}_{\gamma}$ couplings even in a region of small values for $q\bar q\to\gamma\gamma\gamma$ subprocess. The triphoton process is not sensitive to variations of $\bar{c}_{HW}$, $\bar{c}_{HB}$, $\tilde{c}_{HW}$ and $\tilde{c}_{HB}$ couplings,  but only to $\bar{c}_{\gamma}$ and $\tilde{c}_{\gamma}$. Therefore, we will only consider these couplings in the detailed analysis including detector effects through triphoton production at FCC-hh with 100 TeV center of mass energy in the next section.

 \section{Signal and Background Analysis}
 We perform the detailed analysis of $\bar{c}_{\gamma}$ and $\tilde{c}_{\gamma}$  effective couplings via $pp \to\gamma\gamma\gamma$ process for signal including SM contribution as well as interference between effective couplings and SM contributions ($S+B_{SM}$). We consider the relevant backgrounds;  the same final state of the considered signal process including only SM contribution ($B_{SM}$) and $pp\to \gamma\gamma$+jet SM process in which jet may fake a photon ($B_{\gamma\gamma j}$). 500k events are generated at leading order partonic level in \verb|MadGraph5_aMC@NLO v2.6.3.2| for both SM backgrounds as well as 10 different values of CP-conserving and CP-violating couplings in the range between 0.1 and 0.0007. These events are passed through the Pythia 8 \cite{Sjostrand:2014zea} including initial and final parton shower and the fragmentation of partons into hadron. The detector responses are taken into account with FCC detector card in \verb|Delphes 3.4.1| \cite{deFavereau:2013fsa} package. Details of the FCC detector is given in conceptional design report of FCC-hh \cite{fcchh}. In this card, the photon identification efficiency $\epsilon_{\gamma}$= 95\% for |$\eta$|<2.5, the light jet-to-photon mis-identification probability parameterised by the function $\epsilon_{j \to \gamma}$ = 0.002 exp(-$p_T$ [GeV]/30), the ECAL having an energy resolution around 10\%/$\sqrt{E}$ and the hadron calorimetry around 50\%/$\sqrt{E}$ for single particles are assumed. The minimum photon transverse momentum of photon is set to 0.5 GeV and the
requirement of calorimetry acceptance up to $|\eta|\approx 6$ translates into an inner active radius of only 8 $cm$ at a $z$-distance of 16.6 $m$ from the IP. No pile-up contribution is taken into account in our study. All events are analysed by using the ExRootAnalysis utility \cite{exroot} with ROOT \cite{Brun:1997pa}. 

\begin{figure}
\includegraphics[scale=0.8]{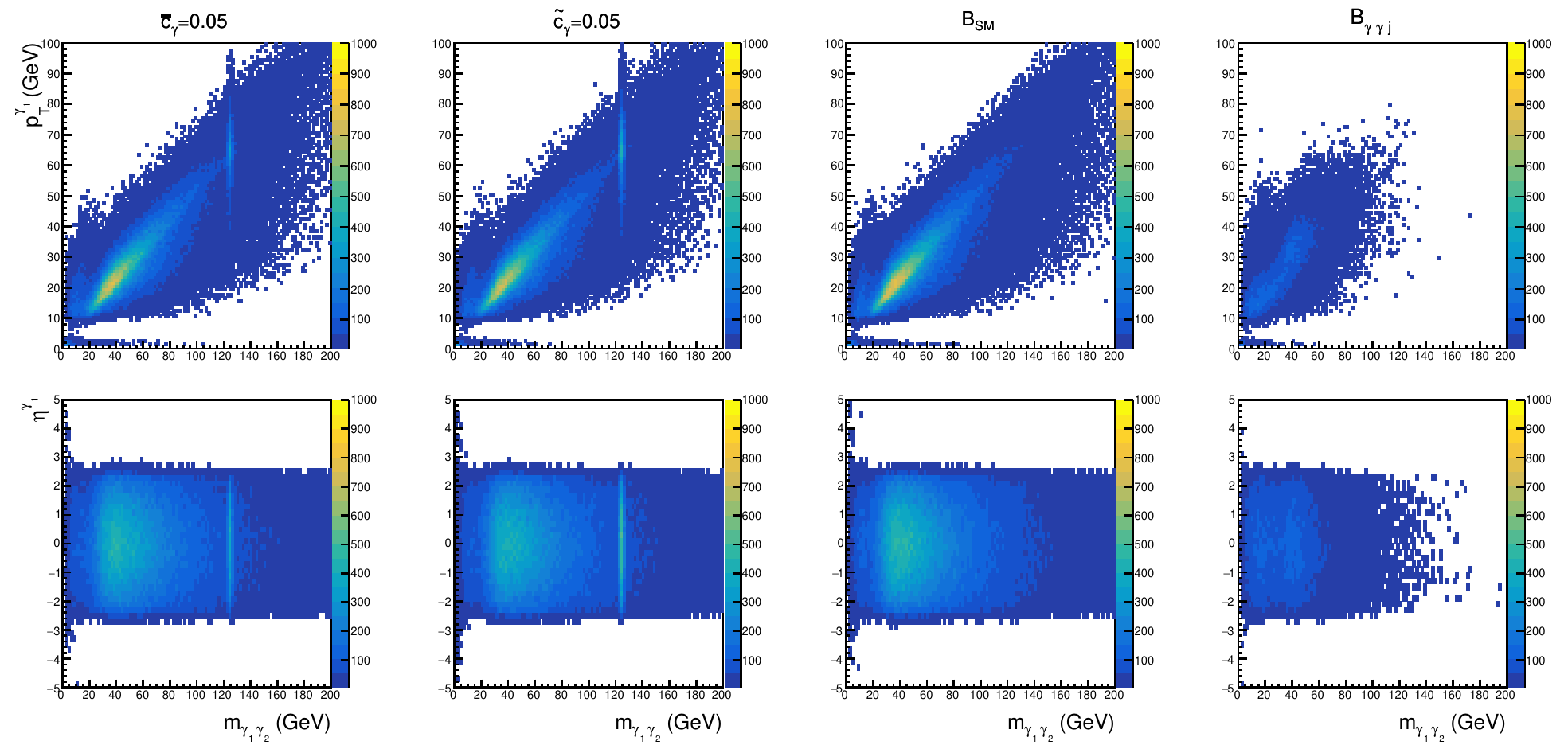} 
\caption{Distributions of transverse momentum (in the first row) and the pesudo-rapidity (in the second row) of the first leading photon versus invariant mass of two leading photons for S+$B_{SM}$ ($\bar{c}_{\gamma} $=0.05) and S+$B_{SM}$ ($\tilde{c}_{\gamma} $=0.05), $B_{SM}$ and $B_{\gamma\gamma j}$  backgrounds (left-to-right).
\label{gamma1}}
\end{figure}

\begin{figure}
\includegraphics[scale=0.8]{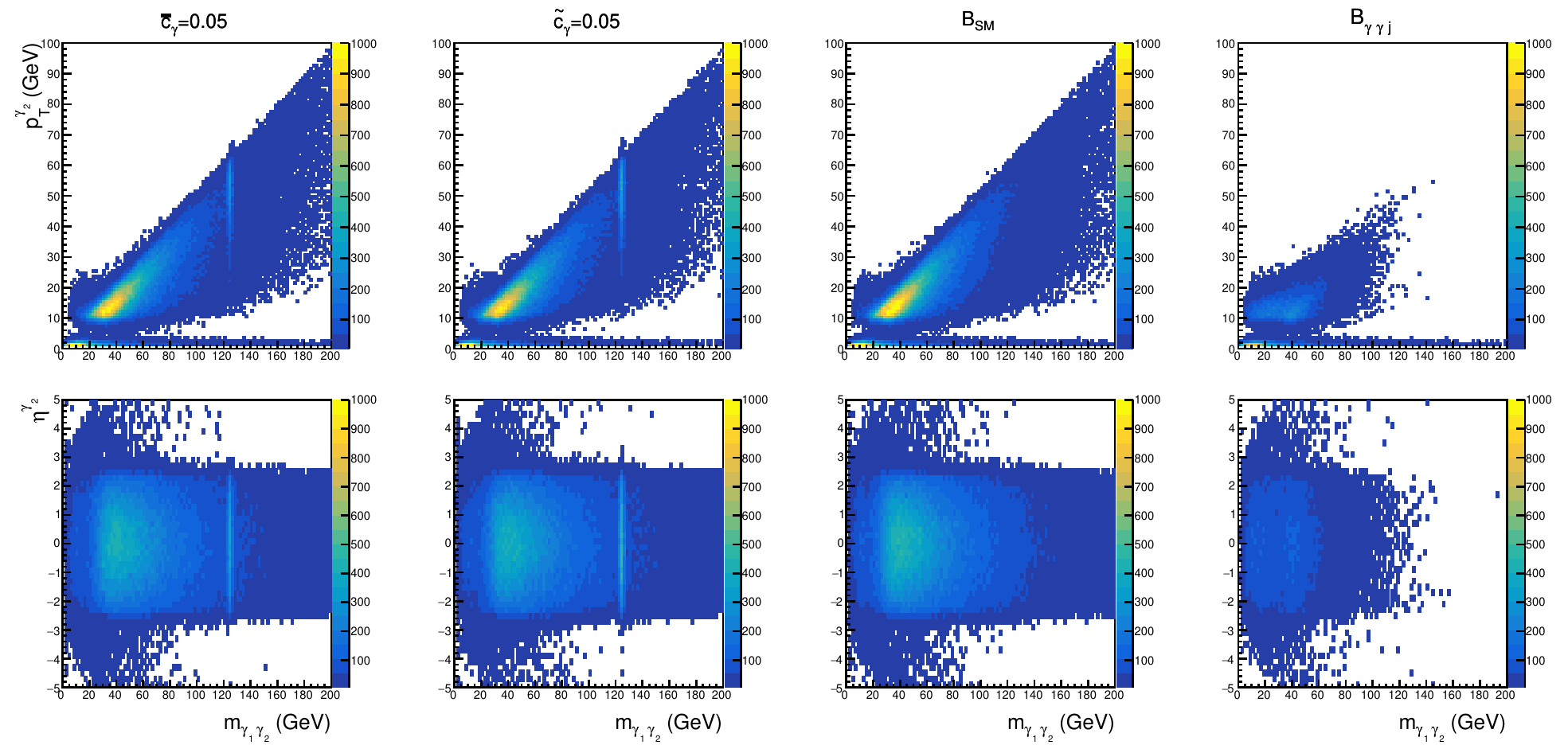} 
\caption{ Distributions of transverse momentum (in the first row) and the pesudo-rapidity (in the second row) of the second leading photon versus invariant mass of two leading photons for S+$B_{SM}$ ($\bar{c}_{\gamma} $=0.05) and S+$B_{SM}$ ($\tilde{c}_{\gamma} $=0.05), $B_{SM}$ and $B_{\gamma\gamma j}$  backgrounds (left-to-right).
\label{gamma2}}
\end{figure}

\begin{figure}
\includegraphics[scale=0.8]{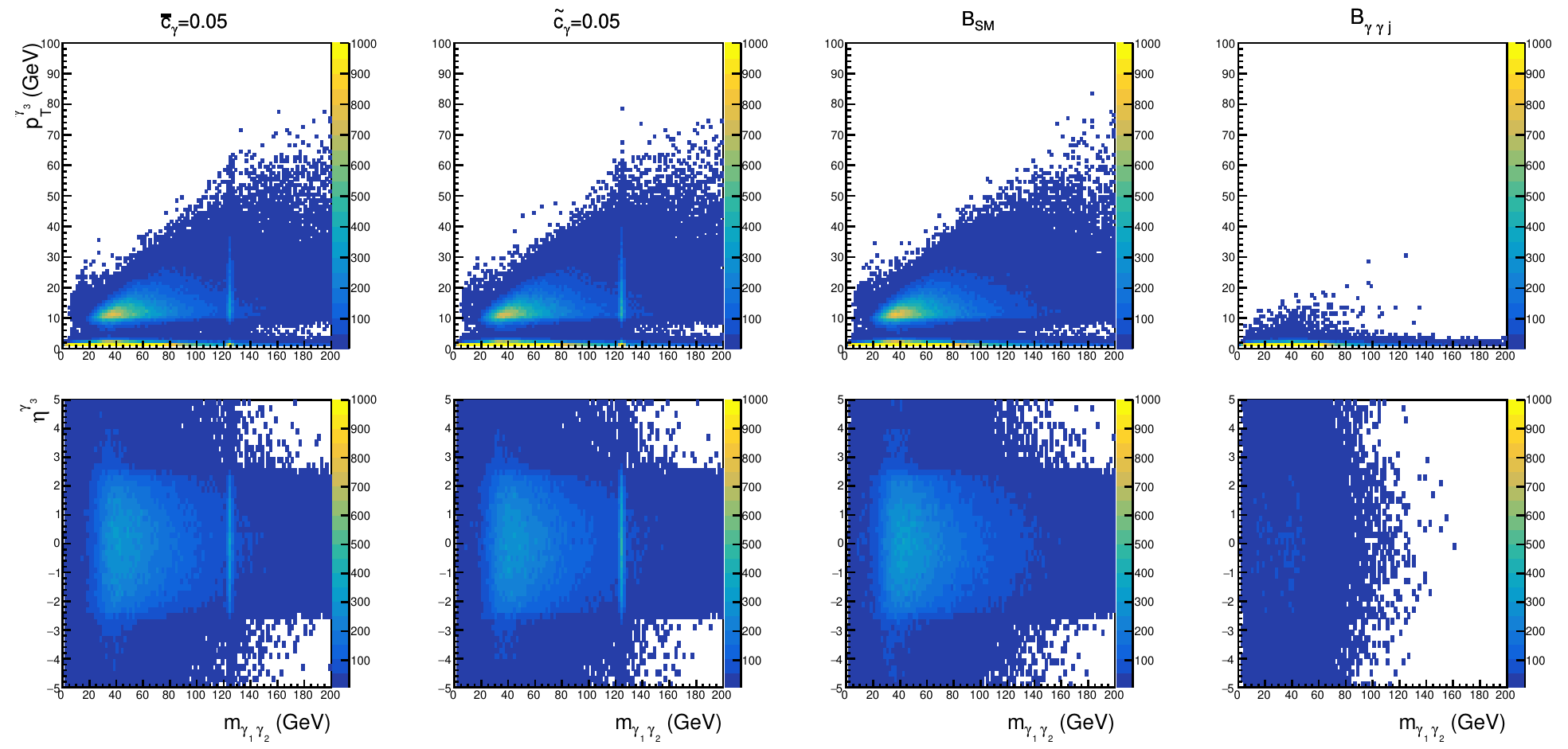} 
\caption{ Distributions of transverse momentum (in the first row) and the pesudo-rapidity (in the second row) of the third leading photon versus invariant mass of two leading photons for S+$B_{SM}$ ($\bar{c}_{\gamma} $=0.05) and S+$B_{SM}$ ($\tilde{c}_{\gamma} $=0.05), $B_{SM}$ and $B_{\gamma\gamma j}$  backgrounds (left-to-right).
\label{gamma3}}
\end{figure}

\begin{figure}
\includegraphics[scale=0.8]{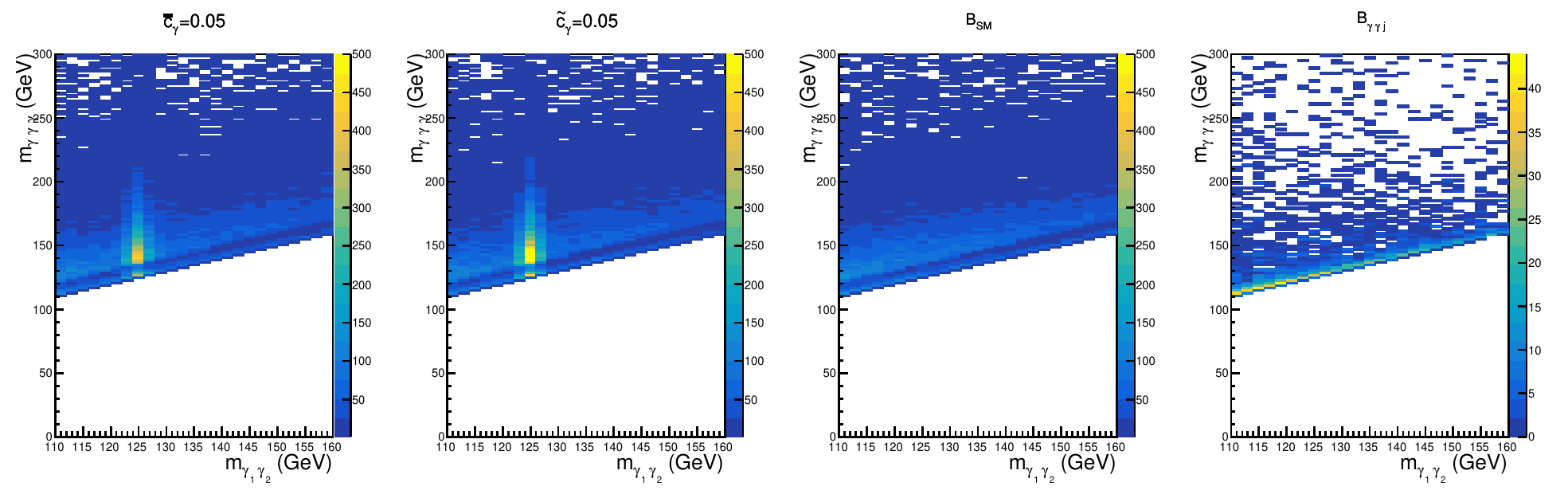} 
\caption{Distribution of invariant mass of three-photons versus invariant mass of two photons for signal $\bar{c}_{\gamma} $=0.05 and $\tilde{c}_{\gamma} $=0.05 and relevant SM Background\label{3gamma}}
\end{figure}

\begin{figure}
\includegraphics[scale=0.4]{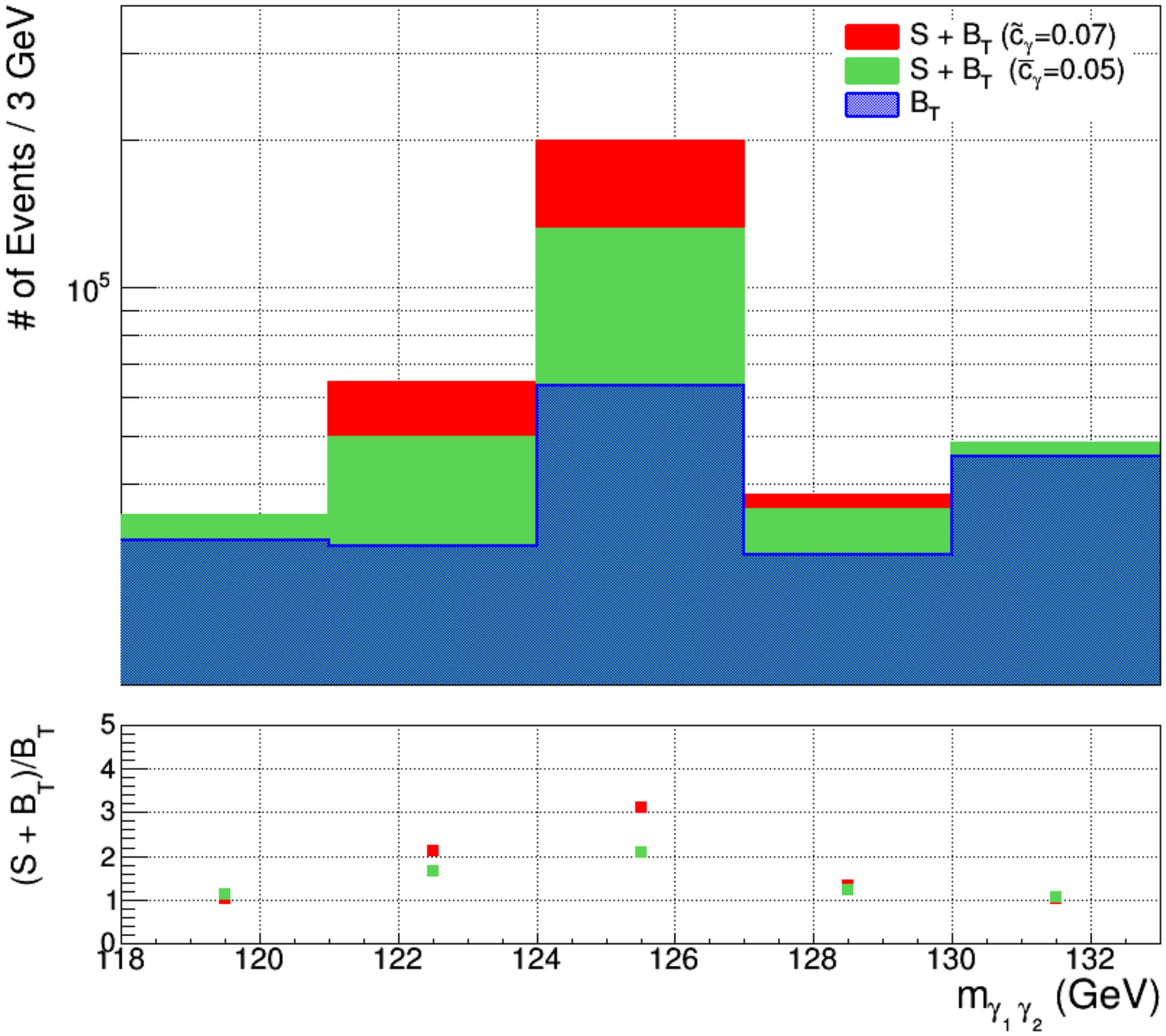} 
\caption{Invariant mass distribution of two photons after all kinematical cuts for signal $S+B_T$ ($\tilde{c}_{\gamma} $=0.07) (red), $S+B_T$ ($\bar{c}_{\gamma} $=0.05) (green) and relevant total SM Background $B_T$ (blue) . \label{final}}
\end{figure}

Requiring at least 3 photons with their transverse momenta ($p_T^{\gamma}$) greater than 0.5 GeV is the pre-selection of the event for detailed analysis. First of all, photons are ordered according to their transverse momentum, i.e., $p_T^{\gamma_1} > p_T^{\gamma_2} > p_T^{\gamma_3}$. In order to obtain the best kinematic cuts to select the signal and background events, transverse momentum ($p_T^{\gamma}$) and pseudo-rapidity ($\eta^{\gamma}$) of the first, second and third leading photons versus the invariant mass of two leading photons for the presence of a signal with values $\bar{c}_{\gamma} $=0.05 and $\tilde{c}_{\gamma} $=0.05 and relevant SM backgrounds are plotted in Fig.~\ref{gamma1}, Fig.~\ref{gamma2} and Fig.~\ref{gamma3} , respectively. There is an apparent lower cut on the $p_T$ around 10 GeV, due to photon trigger set to 0.5 GeV in Delphes card, for invariant masses larger than 10 GeV. These photons are produced by Pythia during initial and final parton shower and the fragmentation of partons into hadron. Comparison of the signal and SM background distributions in Fig.~\ref{gamma1} and Fig.~\ref{gamma2} indicates that $p_T^{\gamma_1}>$ 40 GeV, $p_T^{\gamma_2}>$ 25 GeV  and $|\eta^{\gamma_{1,2}}| < 2.5$ at the region of Higgs mass. In order the prevent distortion of the low end of the invariant mass spectrum of two photon, we use the thresholds in $p_T/m_{\gamma_1\gamma_2}$ rather than fixed cut in  $p_T$. Therefore, we apply  $p_T^{{\gamma_1}({\gamma_2})}/m_{\gamma_1\gamma_2}$ to be greater than 1/3 (1/4) in addition to fixed cut on the transverse momentum of the third leading photon $p_T^{\gamma_3}$ > 12 GeV as seen in Fig.~\ref{gamma3}. Since the photon isolation is a useful requirement to select prompt photons, the minimum distance between each photon is required to satisfy $\Delta R(\gamma_i,\gamma_j)= \left[(\Delta\phi_{\gamma_i,\gamma_j}])^2+(\Delta\eta_{\gamma_i,\gamma_j}])^2\right]^{1/2} > 0.4$ where $\Delta\phi_{\gamma_i,\gamma_j}$  and $\Delta\eta_{\gamma_i,\gamma_j}$ are the azimuthal angle and the pseudo rapidity difference between any two photons, respectively. The invariant mass of three-photons versus the the invariant mass of two photons for the signal $\bar{c}_{\gamma} $=0.05 and $\tilde{c}_{\gamma} $=0.05 and relevant SM Background are shown in Fig.\ref{3gamma}. We also apply  $m_{\gamma_1\gamma_2\gamma_3} > 120$ GeV to exclude distortion of the low end of the invariant mass spectrum of two photon. After all mentioned kinematic cuts, the reconstructed invariant mass of two leading photons is presented for the signal plus total SM backgrounds ($S+B_T( \bar{c}_{\gamma} =0.05$)) and ($S+B_T(\tilde{c}_{\gamma} =0.07$ )) and relevant total SM backgrounds ($B_T=B_{SM}+B_{\gamma\gamma j}$) in Fig.\ref{final}. Finally, events in which reconstructed invariant mass from two leading photons is in the range of 122 GeV $< m_{\gamma\gamma}< 128 $ GeV are used to obtain limits on the anomalous Higgs effective couplings. A summary of the cuts used in the analysis as well as number of events after each cuts is given in Table \ref{cuts}. As seen from this table, final effect of the all cuts is approximately 2.5\%-5.4\% for S+$B_{SM}$($\bar{c}_{\gamma}$=0.05) and S+$B_{SM}$($\tilde{c}_{\gamma} $=0.07) while 0.7\%-0.005\% for other relevant backgrounds, respectively.
\begin{table}
\caption{Number of signal and background events after each kinematic cuts used in the analysis with $L_{int}=10$ ab$^{-1}$. \label{cuts}}
\begin{ruledtabular}
\begin{tabular}{llcccc}
 Cuts &S+$B_{SM}$($\bar{c}_{\gamma} $=0.05)&S+$B_{SM}$($\tilde{c}_{\gamma} $=0.07)&$B_{SM}$&$B_{\gamma\gamma j}$ \\ \hline
Pre-selection ( $N_\gamma ,\geqslant 3$ $N_{jet}=0$)&4489272&3701687&3329733& 1268275000  \\
$p_T^{{\gamma_1}({\gamma_2})}/m_{\gamma_1\gamma_2} >1/3 (1/4)$&1695195&1459231&1225650&4373358\\
$|\eta^{\gamma_{1,2}}| < 2.5$&1651206&1417996&1192817&3699204  \\
$\Delta R(\gamma_i,\gamma_j) > 0.4$&1638638&1407484&1183304&3578202\\
$m_{\gamma_1\gamma_2\gamma_3} > 120$ GeV&754423.5&720973.9&502006.7&916158  \\
122 GeV $< m_{\gamma\gamma}< 128 $ GeV&113170.5&198460.2&23371.5&69144 \\
\end{tabular}
\end{ruledtabular}
\end{table}

The sensitivity of the dimension-6 Higgs-gauge boson couplings in $pp\to\gamma\gamma\gamma$ process is evaluated by means of a $\chi^{2}$ fit to the simulated data. The $\chi^{2}$ function with and without a systematic error is defined as follows
\begin{eqnarray}
\chi^{2} =\sum_i^{n_{bins}}\left(\frac{N_{i}^{NP}-N_{i}^{B}}{N_{i}^{B}\Delta_i}\right)^{2}
\end{eqnarray}
where $N_i^{NP}$ is the total number of events in the existence of effective couplings ($S$) , $N_i^B$ is number of events of relevant SM backgrounds in $i$th bin of the invariant mass distributions of reconstructed Higgs boson from two leading photon, $\Delta_i=\sqrt{\delta_{sys}^2+\frac{1}{N_i^B}}$ is the combined systematic ($\delta_{sys}$) and statistical errors in each bin. In this analysis, we focused on $\bar{c}_{\gamma} $ and $\tilde{c}_{\gamma} $ couplings which are the main coefficients contributing to $pp\to\gamma\gamma\gamma$ signal process. 

\begin{figure}
\includegraphics[scale=0.6]{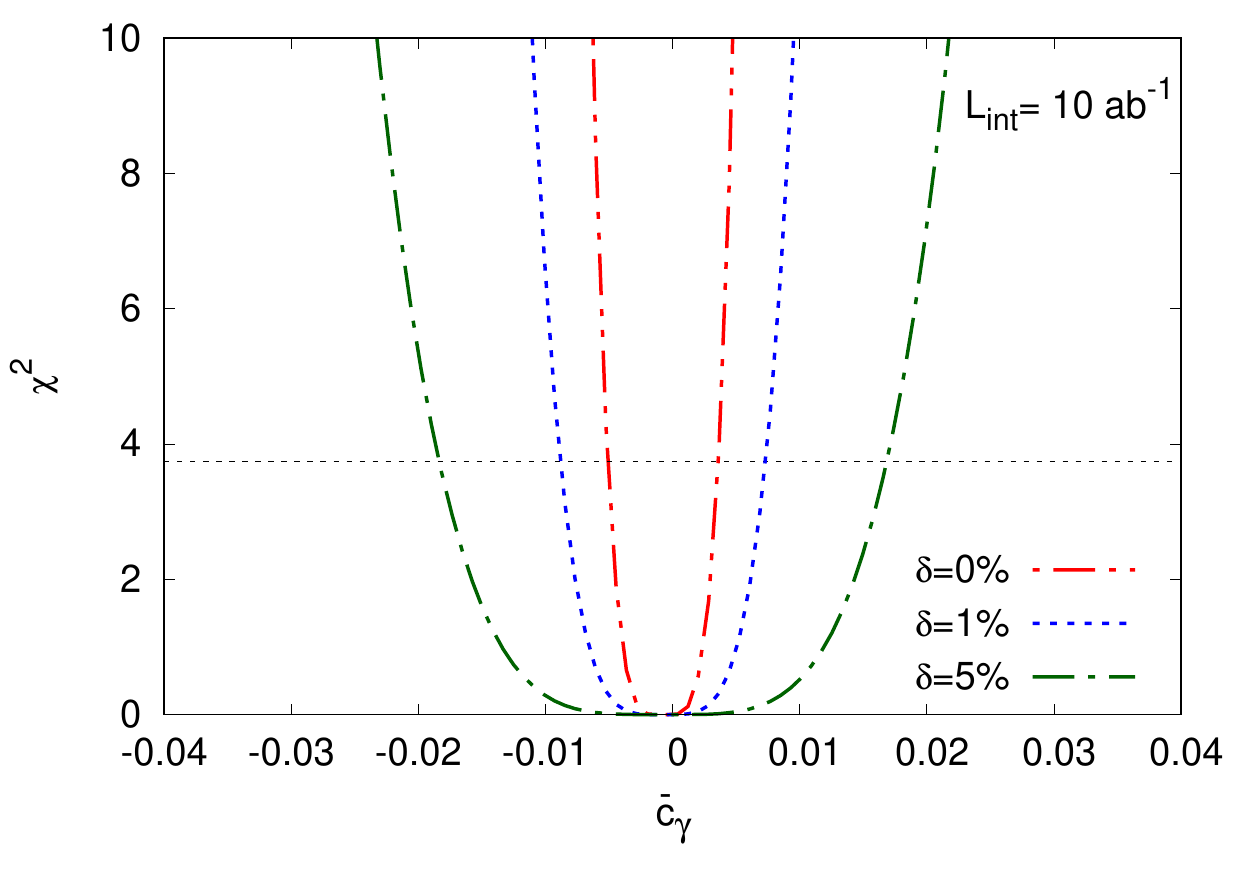} \includegraphics[scale=0.6]{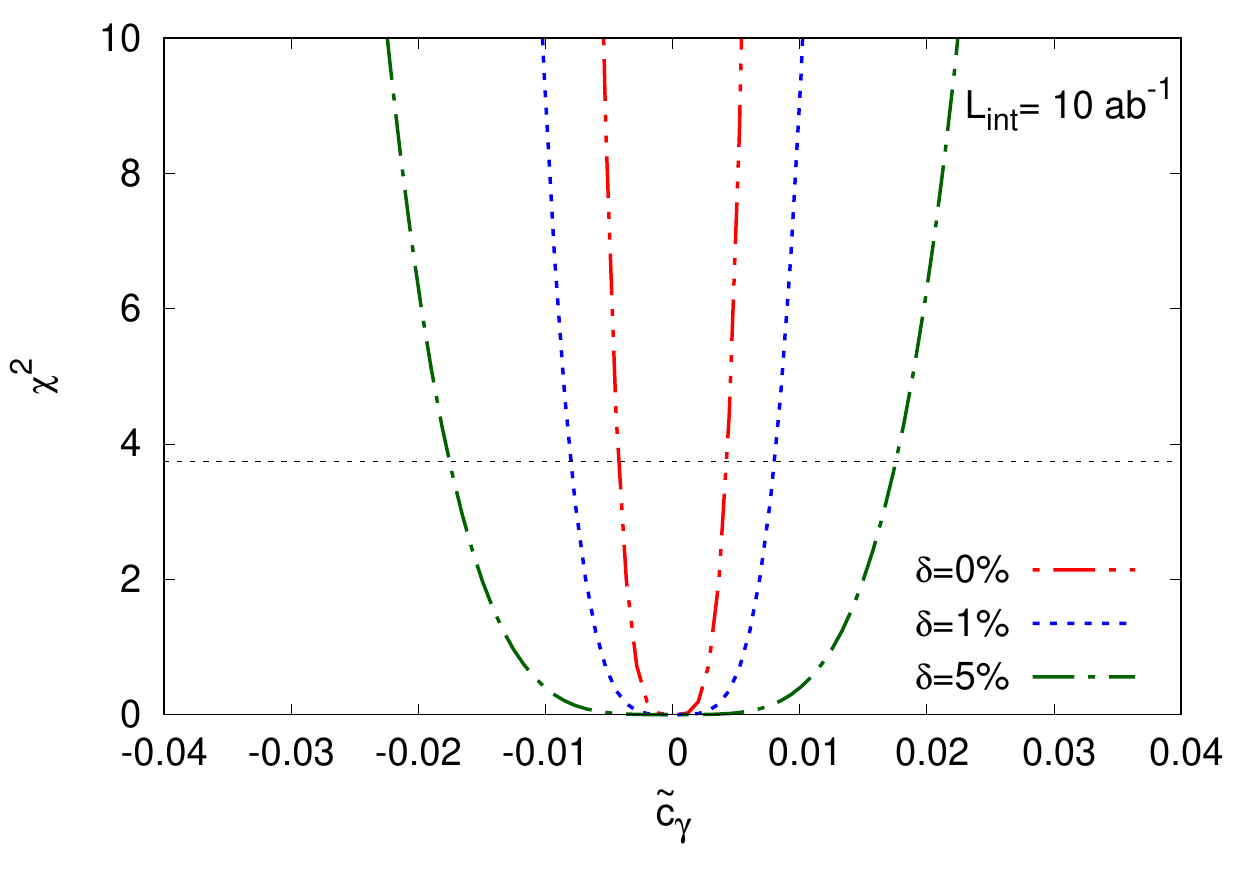} 
\caption{Obtained $\chi^2$ as a functions of $\bar{c}_{\gamma}$ (left) and $\tilde{c}_{\gamma}$ (right) couplings for 100 TeV center of mass energy for the integrated luminosity of 10 ab$^{-1}$ (the dotted line corresponds to 95\% C.L.) including without and with systematic errors (1\% and 5\%). The limits are each derived with all other coefficients set to zero.  \label{limits}}
\end{figure}

 Fig.~\ref{limits} shows the obtained $\chi^2$ value as a function of $\bar{c}_{\gamma}$ and $\tilde{c}_{\gamma}$ couplings for 100 TeV center of mass energy with an integrated luminosity of  $L_{int}=10$ ab$^{-1}$ without and with systematic errors (1\% and 5\%). The 95\% Confidence Level (C.L.) limits without systematic error on dimension-6 Higgs-gauge boson couplings $\bar{c}_{\gamma} $ and $\tilde{c}_{\gamma}$ are [-0.0051; 0.0038]  and [-0.0043; 0.0043], respectively. If  taking into account 5\% systematic error, the obtained limits for Lint=10 ab$^{-1}$, about three times worse than obtained limits without systematic errors. The current experimental limits on these couplings probed using a fit to five differential cross sections measured by ATLAS experiment in $H\to\gamma\gamma$ decay channel with an integrated luminosity of 20.3 fb$^{-1}$ at $\sqrt s$=8 TeV are [-0.00074; 0.0057] and [-0.0018; 0.0018] in Ref. \cite{Aad:2015tna}. However, a similar analysis carried out by ATLAS Collaboration using 36.1 fb$^{-1}$ of proton-proton collision at $\sqrt s$ = 13 TeV did not consider $\bar{c}_{\gamma} $ and $\tilde{c}_{\gamma}$ couplings due to the lack of sensitivity of the $H\to\gamma\gamma$ decay channel \cite{Aaboud:2018xdt}. The high luminosity LHC constraint on CP-conserving coupling $\bar{c}_{\gamma}$ extrapolated from LHC Run1 data with $pp\to H+j$, $pp\to H+2j$, $pp\to H$, $pp\to W+H$, $pp\to Z+H$ and $pp\to t \bar t+H$  production modes using a shape analysis on the Higgs transverse momentum is obtained [-0.00016; 0.00013] at 95 \% CL at the center-of-mass energy of 14 TeV with 3000 fb$^{-1}$ \cite{Englert:2015hrx}. Using Run-1 data with variety of Higgs and electroweak boson production channels, constraint on CP-violating $\tilde{c}_{\gamma}$ coupling is [-0.0012; 0.0012] and expected to be a factor of 2 improvement with the high-luminosity LHC prospects \cite{Ferreira:2016jea}. Phenomenological study on CP-conserving the dimension-six operators via $pp \to H+\gamma$ process have been performed considering a fast detector simulation with Delphes at $\sqrt s$=14 TeV \cite{Khanpour:2017inb}. It is found that the limits on coupling $\bar{c}_{\gamma} $ is expected to be [-0.013; 0.023] and  [-0.0042; 0.0075] with the integrated luminosities of 300 fb$^{-1}$ and 3000 fb$^{-1}$, respectively. Both $Z\gamma$ and $\gamma\gamma$ decays of Higgs boson proceed similarly through loop diagrams in the SM. The branching ratios for the Higgs boson decay to $Z\gamma$ is predicted by the SM to be $(1.54\pm 0.09)\times10^{-3}$ at $m_H=125.09$ GeV, which is comparable with the branching ratio of the Higgs boson decay to a photon pair, $(2.27\pm 0.05)\times10^{-3}$ \cite{deFlorian:2016spz}. Using obtained the limits  on  the coupling $\bar{c}_{\gamma} $ ($\tilde{c}_{\gamma}$), we compute these branching ratios as $1.07\times10^{-3}$($12.25\times10^{-3}$) and $0.41\times10^{-3}$ ($0.57\times10^{-3}$), respectively. Our results  are consistent with SM predictions done so far except BR($H\to\gamma\gamma$ )=$12.25\times10^{-3}$ for $\tilde{c}_{\gamma}$=0.0043 due to obtained relatively large coupling value.

\section{Conclusions}
We have investigated the CP-conserving and  CP-violating dimension-6 operators of Higgs boson with other SM gauge boson via $p p\to\gamma\gamma\gamma$ process using an effective Lagrangian approach at FCC-hh ($\sqrt s=100$ TeV, $L_{int}$=10 ab$^{-1}$). We have used leading-order strongly interacting light Higgs basis assuming vanishing tree-level electroweak oblique parameterize and flavor universality of the new physics sector considering realistic detector effect in the analysis. We have shown the 2D plots of kinematic variables, transverse momentum and pseudo-rapidity of each photon and invariant mass distributions of three photon as function of reconstructed invariant mass of two leading photons to determine a cut-based analysis. The reconstructed invariant mass of Higgs-boson from two leading photons is used to obtain limits on the anomalous Higgs effective couplings. We have obtained 95 \% C.L. limits on dimension-six operators analysing invariant mass distributions of two leading photon in $p p\to\gamma\gamma\gamma$ signal process and the relevant SM background. The $p p\to\gamma\gamma\gamma$ process is more sensitive to $\bar{c}_{\gamma}$ and  $\tilde{c}_{\gamma}$ couplings than the other dimension-six couplings. Our results show that FCC-hh with $\sqrt s=100$ TeV, $L_{int}$=10 ab$^{-1}$ will be able to probe the dimension-six couplings of Higgs-gauge boson interactions in $pp\to\gamma\gamma\gamma$ process especially for $\bar{c}_{\gamma}$ and  $\tilde{c}_{\gamma}$ couplings as [-0.0051; 0.0038]  and [-0.0043; 0.0043] without systematic error, respectively. Finally, including all production modes as well as triphoton production in a global fit  to the experimental data would affect the exclusion ranges and may improve the sensitivities.
  
\begin{acknowledgments}
This work was partially supported by Turkish Atomic Energy Authority (TAEK) under the grant No. 2018TAEK(CERN)A5.H6.F2-20. 
\end{acknowledgments}

\newpage




\begin{thebibliography}{99}
\bibitem{Aad:2012tfa} 
  G.~Aad {\it et al.} [ATLAS Collaboration],
  Phys.\ Lett.\ B {\bf 716}, 1 (2012)
  [arXiv:1207.7214 [hep-ex]].
  
\bibitem{Chatrchyan:2012xdj} 
  S.~Chatrchyan {\it et al.} [CMS Collaboration],
  Phys.\ Lett.\ B {\bf 716}, 30 (2012)
  [arXiv:1207.7235 [hep-ex]].
  
 \bibitem{Buchmuller:1985jz} 
  W.~Buchmuller and D.~Wyler,
  Nucl.\ Phys.\ B {\bf 268}, 621 (1986).
  
\bibitem{Grzadkowski:2010es} 
  B.~Grzadkowski, M.~Iskrzynski, M.~Misiak and J.~Rosiek,
  JHEP {\bf 1010}, 085 (2010)
  [arXiv:1008.4884 [hep-ph]].
  
  \bibitem{Appelquist:1974tg} 
  T.~Appelquist and J.~Carazzone,
  Phys.\ Rev.\ D {\bf 11}, 2856 (1975).
  
\bibitem{Hagiwara:1993qt} 
  K.~Hagiwara, R.~Szalapski and D.~Zeppenfeld,
  Phys.\ Lett.\ B {\bf 318}, 155 (1993)
  [hep-ph/9308347].
  
\bibitem{Corbett:2012ja} 
  T.~Corbett, O.~J.~P.~Eboli, J.~Gonzalez-Fraile and M.~C.~Gonzalez-Garcia,
  Phys.\ Rev.\ D {\bf 87}, 015022 (2013)
  [arXiv:1211.4580 [hep-ph]].
  
\bibitem{Ellis:2014jta} 
  J.~Ellis, V.~Sanz and T.~You,
  JHEP {\bf 1503}, 157 (2015)
  [arXiv:1410.7703 [hep-ph]].
  
\bibitem{Ellis:2014dva} 
  J.~Ellis, V.~Sanz and T.~You,
  JHEP {\bf 1407}, 036 (2014)
  [arXiv:1404.3667 [hep-ph]].

\bibitem{Falkowski:2015fla} 
  A.~Falkowski,
  Pramana {\bf 87}, no. 3, 39 (2016)
  [arXiv:1505.00046 [hep-ph]].

\bibitem{Corbett:2015ksa} 
  T.~Corbett, O.~J.~P.~Eboli, D.~Goncalves, J.~Gonzalez-Fraile, T.~Plehn and M.~Rauch,
  JHEP {\bf 1508}, 156 (2015)
  [arXiv:1505.05516 [hep-ph]].
  
  \bibitem{Ferreira:2016jea} 
  F.~Ferreira, B.~Fuks, V.~Sanz and D.~Sengupta,
  Eur.\ Phys.\ J.\ C {\bf 77}, no. 10, 675 (2017)
  [arXiv:1612.01808 [hep-ph]].
  
\bibitem{Khachatryan:2014kca} 
  V.~Khachatryan {\it et al.} [CMS Collaboration],
  Phys.\ Rev.\ D {\bf 92}, no. 1, 012004 (2015)
  doi:10.1103/PhysRevD.92.012004
  [arXiv:1411.3441 [hep-ex]].
  
\bibitem{Khachatryan:2016tnr} 
  V.~Khachatryan {\it et al.} [CMS Collaboration],
  Phys.\ Lett.\ B {\bf 759}, 672 (2016)
  doi:10.1016/j.physletb.2016.06.004
  [arXiv:1602.04305 [hep-ex]].
  
\bibitem{Aad:2015tna} 
  G.~Aad {\it et al.} [ATLAS Collaboration],
  Phys.\ Lett.\ B {\bf 753}, 69 (2016)
  [arXiv:1508.02507 [hep-ex]].
 
\bibitem{Englert:2015hrx} 
  C.~Englert, R.~Kogler, H.~Schulz and M.~Spannowsky,
  Eur.\ Phys.\ J.\ C {\bf 76}, no. 7, 393 (2016)
  [arXiv:1511.05170 [hep-ph]].

\bibitem{Buckley:2015lku} 
  A.~Buckley, C.~Englert, J.~Ferrando, D.~J.~Miller, L.~Moore, M.~Russell and C.~D.~White,
  JHEP {\bf 1604}, 015 (2016)
  [arXiv:1512.03360 [hep-ph]].

\bibitem{Khanpour:2017inb} 
  H.~Khanpour, S.~Khatibi and M.~Mohammadi Najafabadi,
  Phys.\ Lett.\ B {\bf 773}, 462 (2017)
  [arXiv:1702.05753 [hep-ph]].
  
  \bibitem{Englert:2016hvy} 
  C.~Englert, R.~Rosenfeld, M.~Spannowsky and A.~Tonero,
  EPL {\bf 114}, no. 3, 31001 (2016)
  [arXiv:1603.05304 [hep-ph]].

\bibitem{Degrande:2016dqg} 
  C.~Degrande, B.~Fuks, K.~Mawatari, K.~Mimasu and V.~Sanz,
  Eur.\ Phys.\ J.\ C {\bf 77}, no. 4, 262 (2017)
  [arXiv:1609.04833 [hep-ph]].

\bibitem{Bishara:2016kjn} 
  F.~Bishara, R.~Contino and J.~Rojo,
  Eur.\ Phys.\ J.\ C {\bf 77}, no. 7, 481 (2017)
  [arXiv:1611.03860 [hep-ph]].
  
\bibitem{Liu-Sheng:2017pxk} 
  L.~S.~Ling, R.~Y.~Zhang, W.~G.~Ma, X.~Z.~Li, L.~Guo and S.~M.~Wang,
  Phys.\ Rev.\ D {\bf 96}, no. 5, 055006 (2017)
  [arXiv:1708.04785 [hep-ph]].

\bibitem{Aaboud:2018xdt} 
  M.~Aaboud {\it et al.} [ATLAS Collaboration],
  Phys.\ Rev.\ D {\bf 98}, 052005 (2018)
  [arXiv:1802.04146 [hep-ex]].
  
\bibitem{deCampos:1998xx} 
  F.~de Campos, M.~C.~Gonzalez-Garcia, S.~M.~Lietti, S.~F.~Novaes and R.~Rosenfeld,
  Phys.\ Lett.\ B {\bf 435}, 407 (1998)
  [hep-ph/9806307].
  
\bibitem{GonzalezGarcia:1999fq} 
  M.~C.~Gonzalez-Garcia,
  Int.\ J.\ Mod.\ Phys.\ A {\bf 14}, 3121 (1999)
  [hep-ph/9902321].

\bibitem{Aaboud:2017lxm} 
  M.~Aaboud {\it et al.} [ATLAS Collaboration],
  Phys.\ Lett.\ B {\bf 781}, 55 (2018)
  [arXiv:1712.07291 [hep-ex]].

\bibitem{Achard:2004kn} 
  P.~Achard {\it et al.} [L3 Collaboration],
  Phys.\ Lett.\ B {\bf 589}, 89 (2004)
  [hep-ex/0403037].

  \bibitem{Abreu:1999vt} 
  P.~Abreu {\it et al.} [DELPHI Collaboration],
  Phys.\ Lett.\ B {\bf 458}, 431 (1999).

\bibitem{Heister:2002ub} 
  A.~Heister {\it et al.} [ALEPH Collaboration],
  Phys.\ Lett.\ B {\bf 544}, 16 (2002).


\bibitem{fcc}
Abada, A., Abbrescia, M., AbdusSalam, S.S.{\it et al.} Eur. Phys. J. C  79, 474 (2019)

\bibitem{fcchh}
Future Circular Collider Study. Volume 3: The Hadron Collider (FCC-hh) Conceptual Design Report, preprint edited by M. Benedikt et al. CERN accelerator reports, CERN-ACC-2018-0058, Geneva, December 2018. Submitted to Eur. Phys. J. ST.

\bibitem{Mangano:2017tke} 
  M.~Mangano,
  CERN Yellow Report CERN 2017-003-M
  [arXiv:1710.06353 [hep-ph]].
  
\bibitem{Alloul:2013naa} 
  A.~Alloul, B.~Fuks and V.~Sanz,
  JHEP {\bf 1404}, 110 (2014)
  [arXiv:1310.5150 [hep-ph]].
  
 \bibitem{Contino:2013kra} 
  R.~Contino, M.~Ghezzi, C.~Grojean, M.~Muhlleitner and M.~Spira,
  JHEP {\bf 1307}, 035 (2013)
  [arXiv:1303.3876 [hep-ph]].
  
\bibitem{Alonso:2013hga} 
  R.~Alonso, E.~E.~Jenkins, A.~V.~Manohar and M.~Trott,
  JHEP {\bf 1404}, 159 (2014)
 [arXiv:1312.2014 [hep-ph]].
 
\bibitem{Brivio:2017bnu} 
  I.~Brivio and M.~Trott,
  JHEP {\bf 1707}, 148 (2017)
  [arXiv:1701.06424 [hep-ph]].

\bibitem{Alwall:2014hca}
  J.~Alwall {\it et al.},
  JHEP {\bf 1407} (2014) 079
  [arXiv:1405.0301 [hep-ph]].

\bibitem{Alloul:2013bka} 
  A.~Alloul, N.~D.~Christensen, C.~Degrande, C.~Duhr and B.~Fuks,
  Comput.\ Phys.\ Commun.\  {\bf 185}, 2250 (2014)
 [arXiv:1310.1921 [hep-ph]].

\bibitem{Degrande:2011ua} 
  C.~Degrande, C.~Duhr, B.~Fuks, D.~Grellscheid, O.~Mattelaer and T.~Reiter,
  Comput.\ Phys.\ Commun.\  {\bf 183}, 1201 (2012)
  [arXiv:1108.2040 [hep-ph]].

\bibitem{Sjostrand:2014zea} 
  T.~Sjöstrand {\it et al.},
  Comput.\ Phys.\ Commun.\  {\bf 191}, 159 (2015)
  [arXiv:1410.3012 [hep-ph]].

  \bibitem{deFavereau:2013fsa} 
  J.~de Favereau {\it et al.} [DELPHES 3 Collaboration],
  JHEP {\bf 1402}, 057 (2014)
  [arXiv:1307.6346 [hep-ex]].
  
  \bibitem{exroot}
  http://madgraph.hep.uiuc.edu/Downloads/ExRootAnalysis

\bibitem{Brun:1997pa} 
  R.~Brun and F.~Rademakers,
  Nucl.\ Instrum.\ Meth.\ A {\bf 389}, 81 (1997).

\bibitem{deFlorian:2016spz} 
  D.~de Florian {\it et al.} [LHC Higgs Cross Section Working Group],
  arXiv:1610.07922 [hep-ph].

\end{thebibliography}
\end{document}